\documentclass{aastex62}
\usepackage{graphicx,graphics,aas_macros,amsmath}
\usepackage{natbib}
\usepackage{booktabs}
\usepackage{color}
\usepackage{amssymb}
\usepackage{mathtools}
\usepackage{xspace}
\usepackage{rotating}
\usepackage{wrapfig}
\usepackage{appendix}
\usepackage{lipsum}

\newcommand{\src}{SMC X-1~}

\newcommand{\e}{\ensuremath{\pm}}
\newcommand{\xmm}{XMM-Newton~}

\newcommand{\suzaku}{\emph{Suzaku}}
\newcommand{\nustar}{\emph{NuSTAR}}
\newcommand{\ecyc}[1]{\ensuremath{E_{\rm{C}}}}

\newcommand{\nhh}{\ensuremath{N_{\rm{H2} }}}
\newcommand{\spo}{super-orbital\xspace}

\begin{document}
\title{Is superorbital modulation in \src caused by absorption in warped precessing accretion disc?}

\author{Pragati Pradhan}\thanks{pragati@mit.edu}
\affiliation{Massachusetts Institute of Technology, Kavli Institute for Astrophysics and Space Research, 70 Vassar St., Cambridge,
MA, 02139, USA}
\author{Chandreyee Maitra}
\affiliation{Max Planck Institute For Extraterrestrial Physics, 85748 Garching, Germany}
\author{Biswajit Paul}
\affiliation{Raman Research Institute, Astronomy and Astrophysics, C. V. Raman Avenue, Bangalore 560080. Karnataka India}

\shorttitle{Superorbital modulation in \src\ }
\shortauthors{P. Pradhan et al.}

\begin{abstract}
We present a broadband spectral-timing analysis of SMC X-1 at different intensity states of its super-orbital variation using 10 \suzaku~and 6 \nustar~observations. The spectrum in all the states can be described by an absorbed powerlaw with a high energy cutoff and a black-body component along with an iron emission line. Compared to other supergiant HMXBs, the Fe K$\alpha$ line equivalent width is low in SMC X-1 - from less than 10 eV in high state to 
upto $\sim$ 270 eV in the low states. 
The spectral shape is dependent on flux with the hard X-ray spectrum steepening with increasing flux. We also report a highly variable normalisation of the power-law component across these 16 \spo states. Pulsations in the hard X-rays for both the instruments were detected in all but two observations. The pulse profiles are near sinusoidal with two peaks and the relative intensity of the second peak decreasing with decreasing luminosity. These findings suggest that the \spo modulation in SMC X-1 is not caused by absorption in  precessing warped accretion disc alone and there are intrinsic changes in X-rays emanating from the neutron star at different \spo states.
We also note a putative cyclotron line at $\sim$ 50 keV in the \nustar~spectra of three bright states indicating a possible magnetic field of $\sim$ 4.2 $\times$ 10$^{12}$ G. Finally, with the new pulse period measurements reported here, the time base for the secular spin-up of \src is increased by thirteen years and the complete pulse period history shows a sudden change in the spin-up trend around 1995.
\end{abstract}

\section{Introduction}

Super-orbital periods are long term periodic/quasi-periodic intensity variations at timescale often several times the orbital period seen in X-ray binaries (an optical companion with a compact object). For X-ray binary systems with an accretion disk (e.g., LMC X-4), this super-orbital modulation is ascribed {\it ad-hoc}~to the presence of a precessing warped accretion disk (PWAD). These warps in the accretion disk may be caused by an interplay of tidal force (from the companion star), viscous drag of different layers of accretion disk and driven by the intense radiation pressure from the compact object \citep[]{larwood1998,ogilvie2001}. As these warps precess with the disk rotation and cause obscuration of the X-rays from the compact object, we see X-ray intensity variations with super-orbital periodicity.

On the other hand, some objects, especially those (but not limited to) that accrete via stellar wind, such \spo modulations are thought to occur due to changes in mass-loss rate, $\Delta\dot{M}$ (which subsequently change the accretion rate on to the compact object). There have been many explanations as to what causes such variable $\Delta\dot{M}$. For example, oscillations in companion stars can cause such variability in $\Delta\dot{M}$ \citep{Koenigsberger2006} and subsequent variation in X-ray luminosity - provided there is a mechanism to keep such oscillations stable \citep{farrell2008}. Other possibilities for \spo modulation include the existence of corotation interaction regions (CIRs, e.g., in IGR J16493-4348; \citealt{bozzo2017cir}), accretion bulge formed by collision of the stellar wind with the outer edge of the PWAD \citep{Zdziarski2009}, or the formation of a transient disk (e.g., in 4U 0114+65; \citealt{hu2017}), magnetic axis precession (where compact object is a neutron star, e.g., Her X-1; \citealt{Postnov2013}), presence of a third body, or even jet formation (e.g., SS 433, \citealt{margon1984}). Furthermore, new discoveries of \spo modulation from in all kinds of accreting systems (Supergiant Fast X-ray Transients; SFXTS, classical supergiant X-ray binaries; SgXBs, black hole; BH, Ultra-Luminous X-rays; ULXs) independent of the inclination angles indicate that this effect of \spo modulation is not caused by viewing geometry alone \citep{corbet2014}. For an review on this topic,  we refer the reader to \citealt{kotze2012, corbet2014} and references therein.

In this paper, we will investigate the possible cause(s) for \spo modulation in another high-mass X-ray binary, SMC X-1. \src is an eclipsing binary system discovered in 1971 \citep{price1971_smcx1} located in the Small Magellanic Cloud at a distance of $\sim$ 60 kpc \citep{dist_smcx1}. 
The pulsar has a spin period of $\sim$ 0.71 s \citep{lucke1976_smcx1}, an orbital period of $\sim$ 3.9 d with a B0 I supergiant as the companion and exhibit superorbital variability with a periodicity in the range of 40-60 d \citep[]{gruber1984,wojdowski1998}. Given the similarity of \src with other sources like Her X-1 and LMC X-4, the superorbital variation in \src - like these two sources - was ascribed to the PWAD \citep{wojdowski1998}. Unlike these two sources however, the superorbital period in \src is not periodic but varies within 40-60\,d repeating cyclically after $\sim$ 7 years each \citep{clarkson2003_smcx1}. Such a variability in the \spo period for \src can theoretically be attributed to the presence of different warping {\it modes} in the accretion disk \citep{ogilvie2001}. 

It is however interesting to note that long-term ASM (1.3-12\,keV) and BATSE (20-100\,keV) light-curves of \src covering very different energies show similar variations in the \spo modulation (Fig.~1 of \citealt{clarkson2003_smcx1}). This similarity in the \spo modulation across the entire energy range of 1.3-100\,keV cannot be explained if the \spo modulation in \src is a result of absorption by neutral matter in the PWAD (since soft X-rays are absorbed more than hard X-rays). Such lack of absorption signatures was also reported in the original paper by \citet{wojdowski1998}. Later authors also argue that such a complex \spo behaviour in \src cannot be understood by assuming simple models of PWAD alone (see Discussions of \citealt{clarkson2003_smcx1,Trowbridge2007}).

In this paper, we investigate the broad band X-ray spectral and timing characteristics of \src at different superorbital phases to examine the proposed scenario of intensity variation by absorption in the precessing warped disk. The X-ray spectrum of \src is usually described with two components: a hard power-law component and a soft-thermal component with the latter possibly arising from the reprocessing of soft X-rays \citep{paul2002_smcx1}. A remarkable dissimilarity and phase lag of the soft X-ray with the hard X-ray pulse profiles has also been noted in literature \citep{paul2002_smcx1,neilsen2004,hickox2005}. \\
The X-ray spectrum of neutron star High Mass X-ray binaries (HMXBs) also usually feature iron K$\alpha$ emission lines which are produced by fluorescence emission of either neutral or partially ionized matter in the accretion disk or circumstellar matter around the neutron star. One interesting aspect of \src is its remarkably weak iron line in the X-ray spectrum compared 
to other supergiant HMXBs \citep{garcia2015}. \\
In this paper, we present, for the first time, a comprehensive view of the different spectral states of the super-orbital variation of \src through simultaneous broadband spectral fitting (and timing analysis) using \suzaku~and \nustar~observations. In Section \ref{sec:obs}, we present the details of the observations and data reduction. In section \ref{sec:analysis and results}, we present the details of data analysis and results followed by discussion in section \ref{sec:disc} and finally a summary in section \ref{sec: conc}.
\vspace{0.6cm}
\section{Observations and data reduction}
\label{sec:obs}
\subsection{\suzaku}
\src was observed with the \emph{Suzaku} observatory \citep{M07} ten times during 2011-2012. \emph{Suzaku} consists of two main payloads: the X-ray Imaging Spectrometer (XIS, 0.2-12 keV; \citealt{K07}) and the Hard X-ray Detector (HXD, 10-600 keV; \citealt{T07}). The XIS consists of four CCD detectors of which three (XIS 0, 2 and 3) are front illuminated 
(FI) and one (XIS 1) is back illuminated (BI). The HXD comprises PIN diodes and GSO crystal scintillator detectors. \\
The data reduction was done on the filtered `cleaned' event files following the 
reduction technique mentioned in the same Suzaku ABC guide\footnote{http://heasarc.gsfc.nasa.gov/docs/suzaku/analysis/abc/}. We applied the barycentric correction to all event files using \texttt{aepipeline}. 
In case of CCD data as obtained by XIS, we had to investigate the effect of pile-up which is two photons of lower energy being read as one with higher energy, thereby causing artificial hardening of the X-ray spectrum. Therefore, for those observations affected by pile-up, we discarded 
photons collected within the portion of the PSF where the estimated pile-up fraction was greater than 4 \% determined using the FTOOLS task \texttt{pileest}. 
XIS lightcurves and spectra were then extracted by choosing circular regions of $\sim$ $3^{'}$, or $4^{'}$ radius from the source position depending on whether the observation was made in 1/4 or 0 window mode, respectively.
Background for the XIS were extracted by selecting regions of the same size as mentioned above in a portion of the CCD that was not significantly contaminated by the 
source X-ray mission. \\
Being a photon counting detector, data from PIN detector have to be corrected for deadtime which is the time interval for which the detector 
electronics are processing one photon and thus cannot yet detect the arrival of another. This dead time correction was done using \texttt{FTOOLS} task \texttt{hxddtcor}. 
For the HXD/PIN, simulated `tuned' non X-ray background event files (NXB) corresponding to the month and year of the respective observations 
were used to estimate the non X-ray background \footnote{http://heasarc.nasa.gov/docs/suzaku/analysis/pinbgd.html}\citep{F09}. \\
The XIS spectra were extracted with 2048 channels and the PIN spectra with 255 channels. 
Response files for the XIS were created using the CALDB version `20150312'. For the HXD/PIN spectrum, response files corresponding to the epoch 
of the observation were obtained from the \emph{Suzaku} guest observer
facility\footnote{http://heasarc.nasa.gov/docs/heasarc/caldb/suzaku/}. 
The observation details are presented in Table \ref{obsid}. For brevity, in this paper, we denote each observation by the last two numbers of its OBSID (eg, 706030010 is denoted by 10), except 706030100 which is abbreviated as 100. The observations 10, 20, 50, 70, 80,90 were in high states (H), 30, 100 in medium (M) and 40, 60 in low states (L). 

\subsection{\nustar}
\src was observed with \nustar~(\emph{The Nuclear Spectroscopic Telescope Array} ; \citealt{Harrison_2013}) six times during 2012-2016 at various superorbital phases (reported in Table \ref{obsid} ). \nustar~consists of two focal plane modules, FPMA and FPMB, each made up of four pixelated detectors (DET0-DET3) spanning an energy range of 3-79\, keV. We used \texttt{nupipeline} version 0.4.6 to generate cleaned event files which also provides the recommended source and back ground regions\footnote{https://heasarc.gsfc.nasa.gov/docs/nustar/analysis/}. We extracted the source spectrum and background spectrum using these corresponding region files. Further, we filtered for good time intervals (GTIs) for the duration of the simultaneous observations in FPMA and FPMB.  

Finally, following the same naming convention as \suzaku~, we short-hand the six \nustar~observations by the last two digits of their OBSID. For example, 30202004008 is denoted as 08, 30202004002 as 02 and so on. The observations 08, 02, 03 were in high state (H), 04 in medium (M) and 01, 06 in low state (L). 
\vspace{0.6cm}
\section{analysis and results}
\label{sec:analysis and results}
\subsection{Timing analysis}
\begin{figure}[h]
\begin{center}
        \begin{minipage}{0.9\textwidth}
    \includegraphics[height=7cm,width=13.8cm]{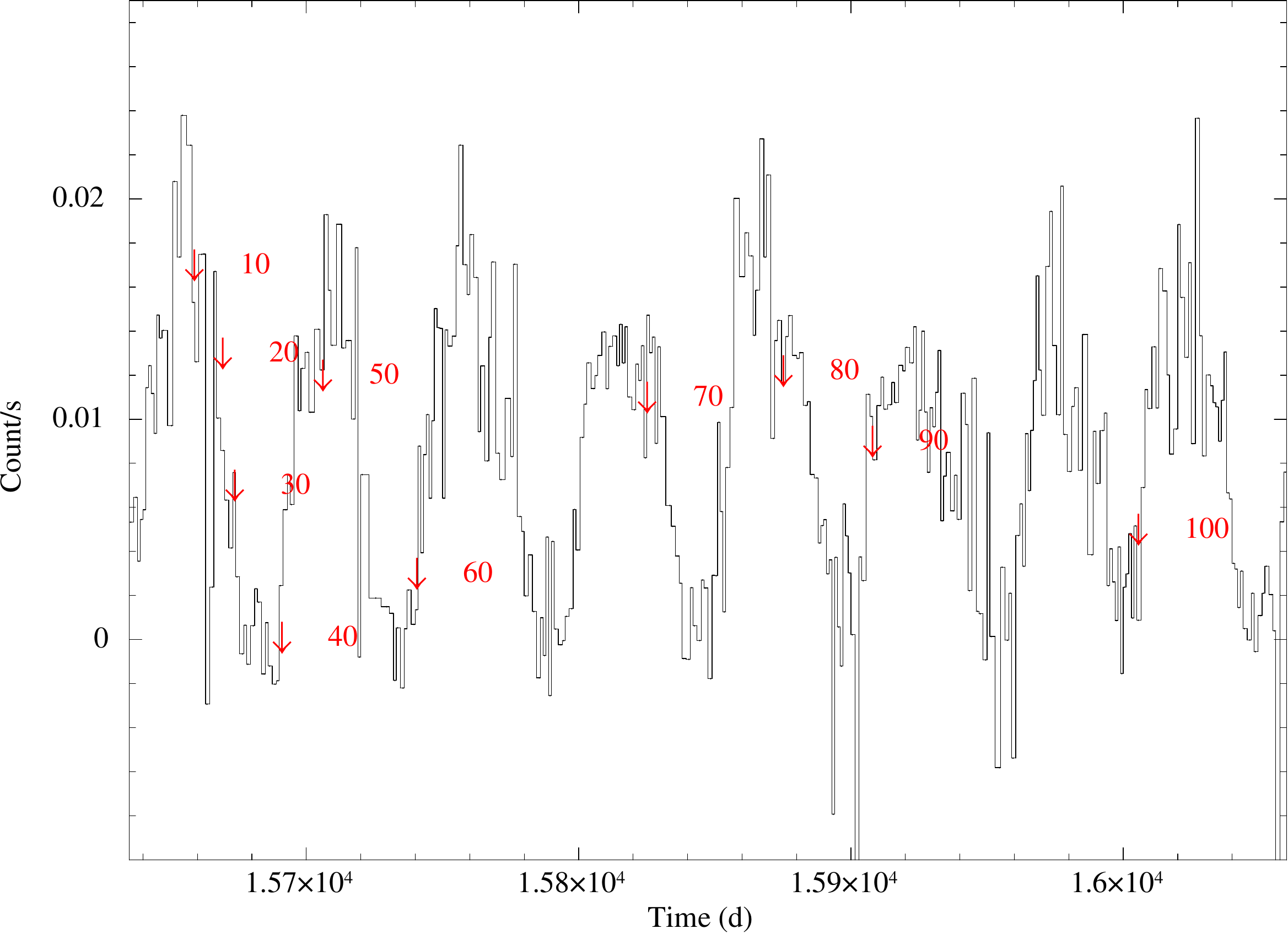}
    \vspace{10pt}
    \includegraphics[height=7cm,width=14cm]{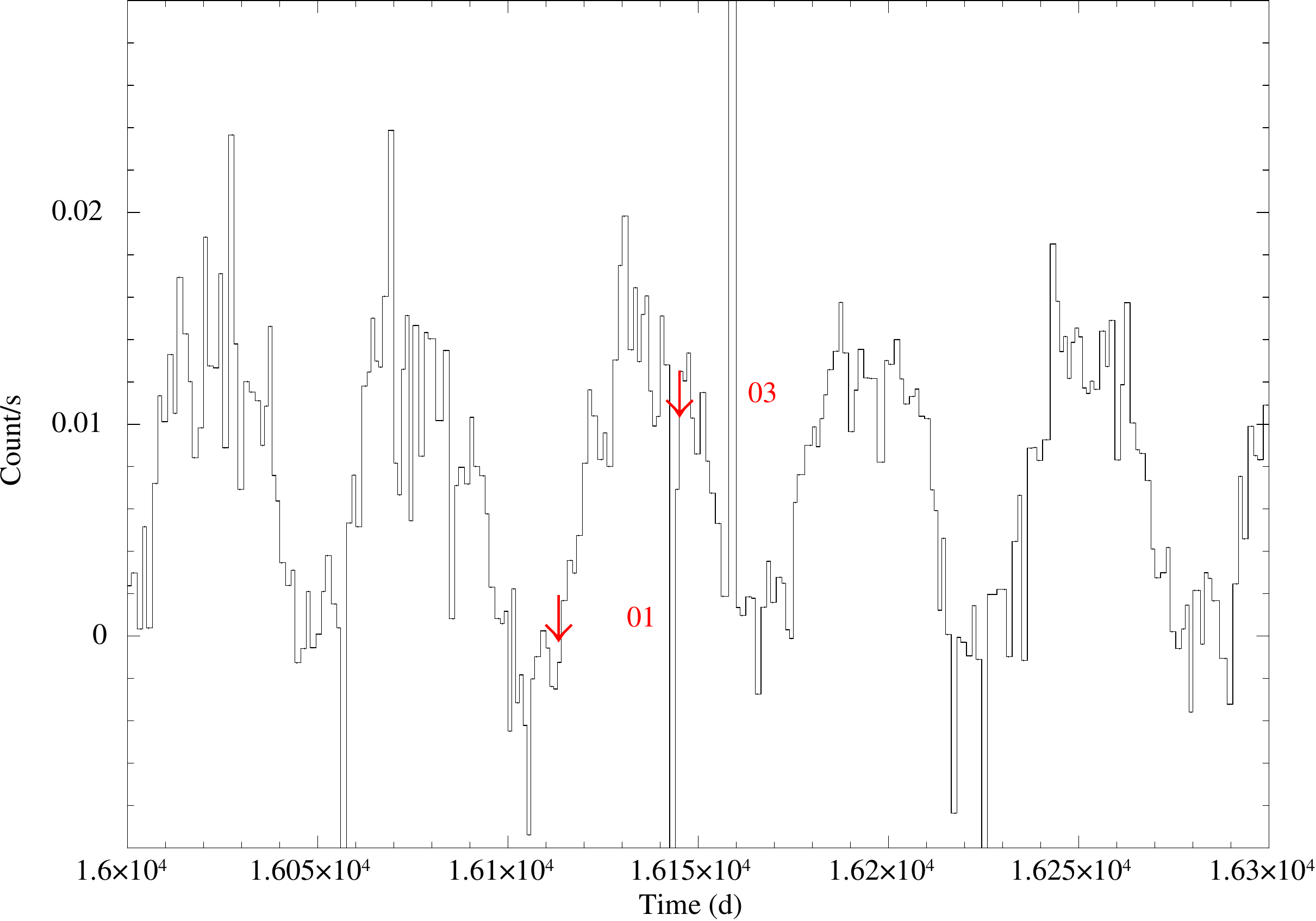}
    \vspace{10pt}
    \includegraphics[height=7cm,width=14cm]{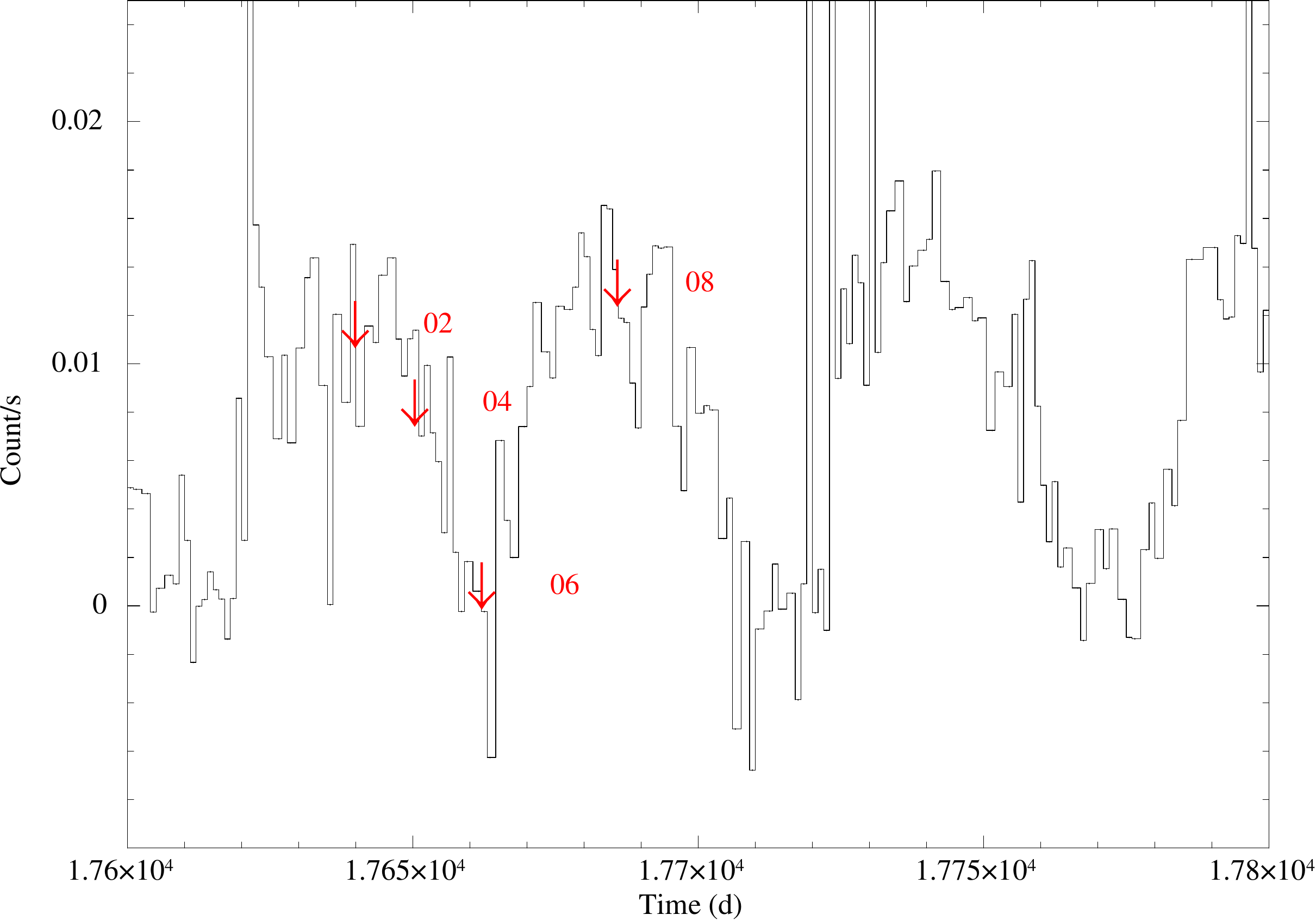}
\end{minipage}
\caption{One day binned Swift/BAT lightcurve of \src filtered for eclipse and ingress/egresses. Each number in the lightcurve stands 
for different OBSID of \suzaku~(top) and \nustar~(middle and below). The figures are marked with aliases of OBSIDs as outlined in Table \ref{obsid}. The errors have been removed for plotting to allow the labelling to be clearly seen. }\label{markings_bat1}
\end{center}
\end{figure}

In order to get an overview of the long-term superorbital intensity variation in these 16 observations, we plotted the one-day binned Swift/BAT lightcurve filtered for eclipses and ingresses/egresses of the NS. This is shown in Fig.~\ref{markings_bat1} where we also mark the \suzaku~and \nustar~observations. Note that we removed the error bars in these lightcurves to assist visual clarity.

\label{timing analysis}
\subsubsection{\emph{\suzaku}~timing analysis}
For timing analysis with \emph{Suzaku} data, we extracted background subtracted light curves from the available XIS and PIN barycentric corrected event files. Since the spin period of \src is $\sim$ 0.7 s and the lowest available observation mode for XIS was 1/4 window that collect data at 2\,s interval, we could not use the XIS lightcurves to detect pulsations.
The PIN lightcurves were background subtracted by generating a background lightcurve using the simulated background files\footnote{http://heasarc.nasa.gov/docs/suzaku/analysis/pinbgd.html}\citep{F09}. We use the background corrected PIN data extracted with a resolution of 0.01 s to search for periodicity in the individual observations and creation of lightcurves after relevant orbital corrections. In order to correct for Doppler shift of the pulse period due to the orbital motion of the pulsar, we corrected for arrival time in the PIN lightcurves of individual observations using the orbital parameters from Table 3 of \citet{raichur2010} extrapolated to the time of each observation. 

\subsubsection{\emph{\nustar}~timing analysis}
Similarly, to get a combined lightcurve for each \nustar~observation, we add the individual FPMA and FPMB barycenter-corrected (using \texttt{barycorr}) lightcurves (binned at 0.01 \,s). We correct the arrival times of lightcurves for orbital motion using the orbital parameters from Table 3 of \citet{raichur2010} extrapolated to the time of each observation - similar to that for the \suzaku~lightcurves above.

\subsubsection{Pulse profile and pulse period evolution}
\begin{wrapfigure}{9}{0.5\textwidth}
\centering
\includegraphics[height=8.0cm,width=9.0cm,angle=0]{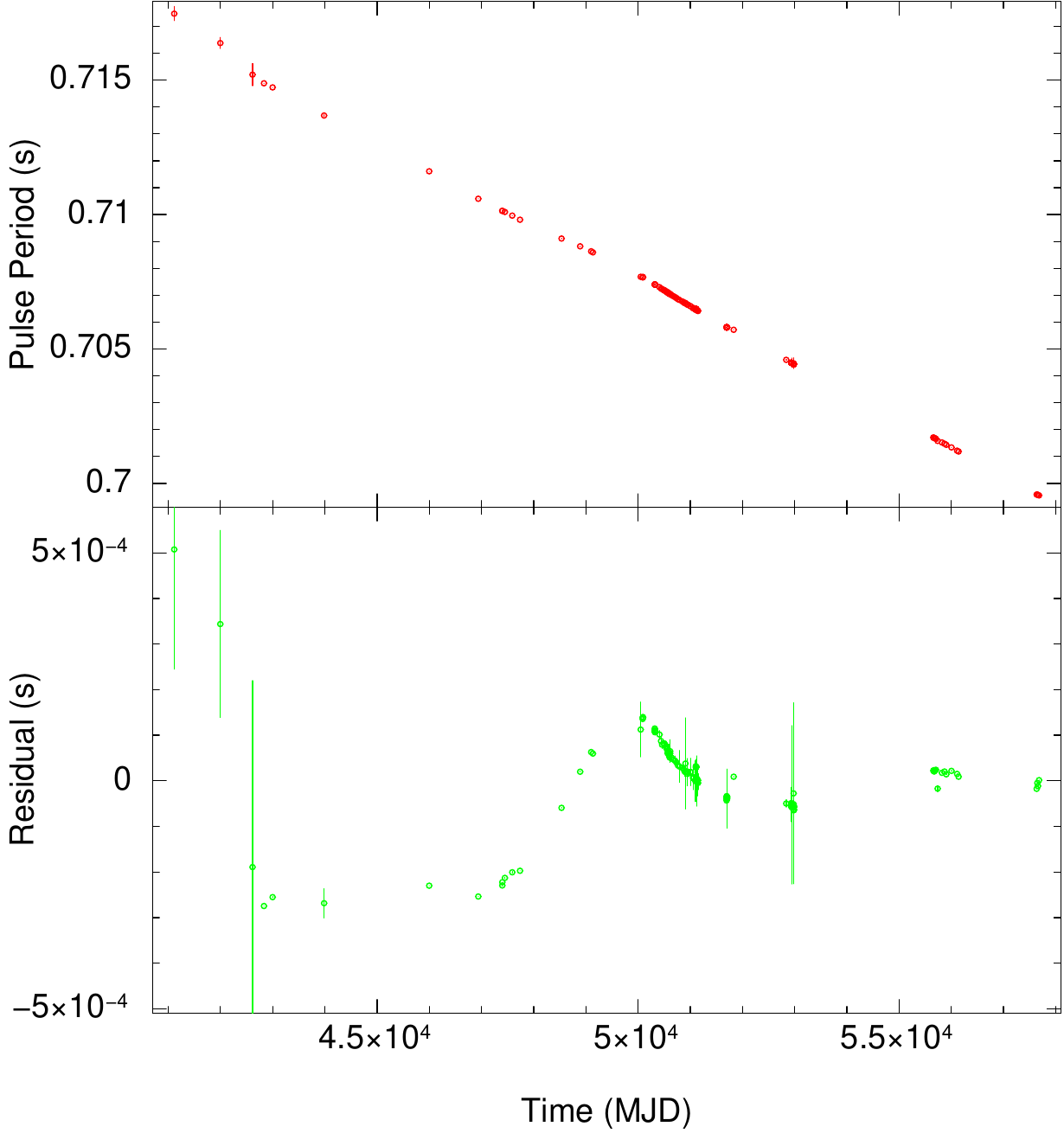}
\caption{\rmfamily{The pulse period history of \src is shown in the top panel. Bottom panel shows the residuals after subtraction of a linear fit. A sudden change in the spin-up rate around MJD 50000 is obvious from the residuals.}}
\label{period_evolution}
\vspace{-0.5cm}
\end{wrapfigure}
\begin{figure*}
\includegraphics[height=13.0cm,width=8.0cm,angle=0]{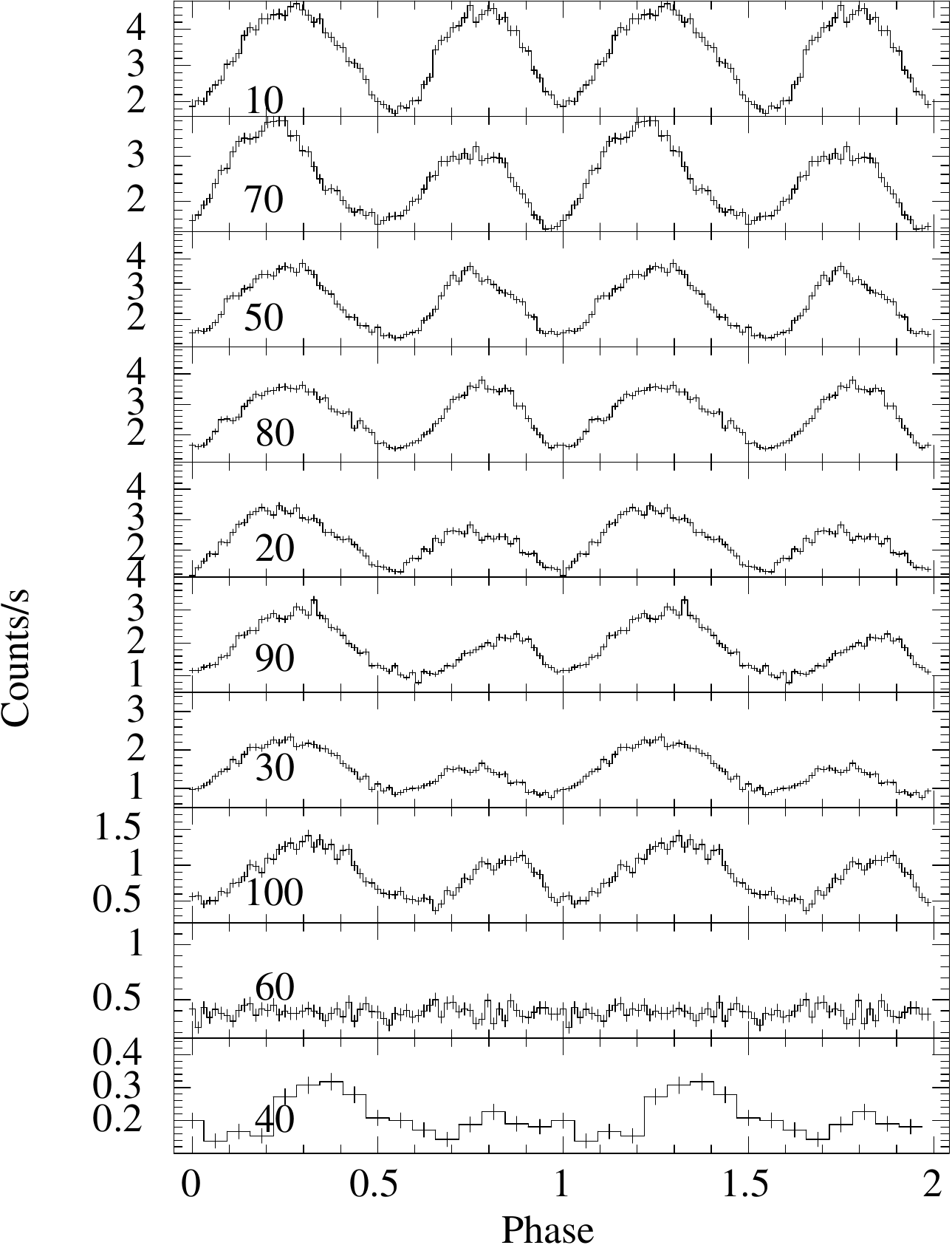}
\includegraphics[height=10.0cm,width=8.0cm,angle=0]{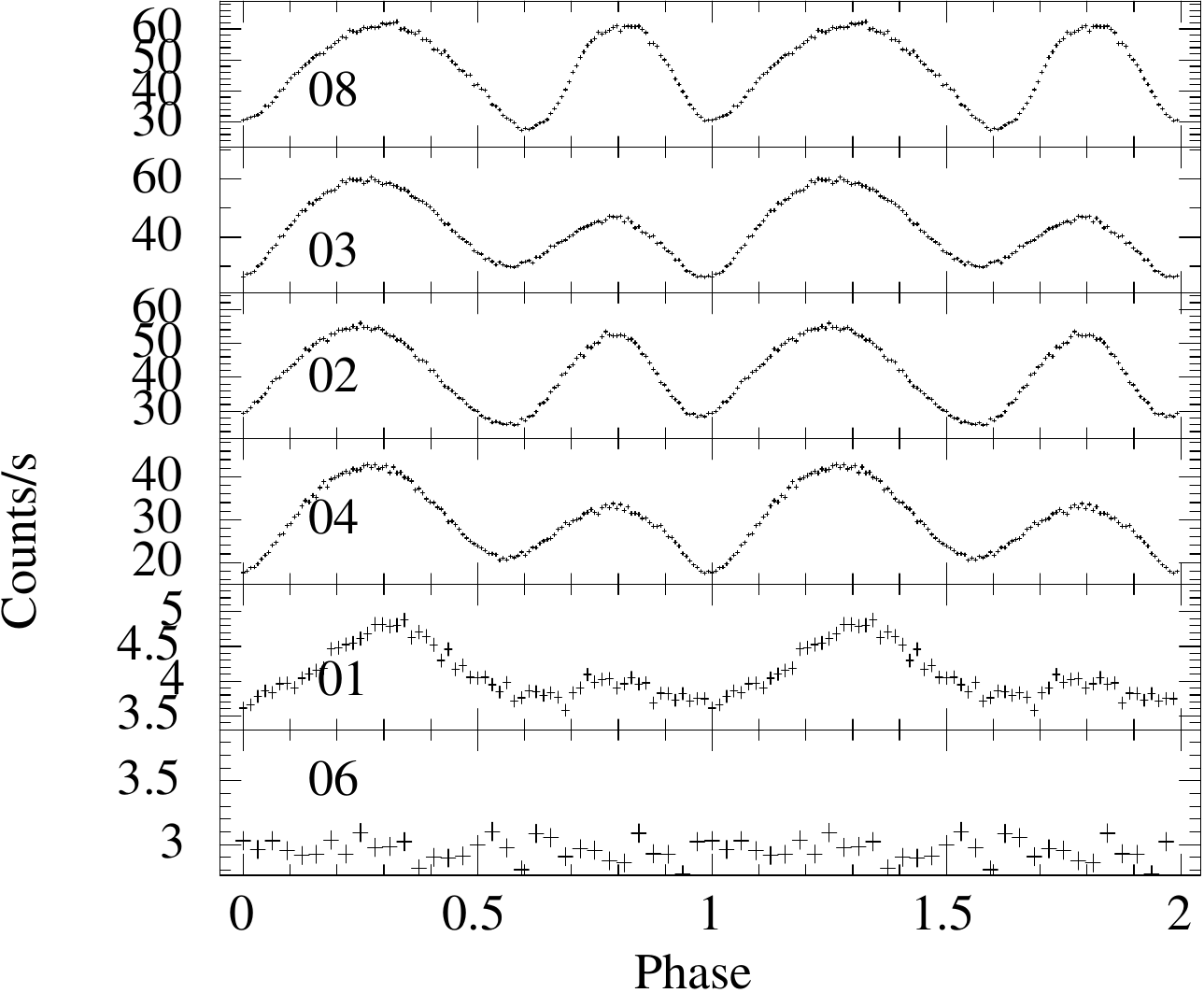}
\caption{\rmfamily{Pulse profiles for all \suzaku~HXD/PIN (left) and \nustar~(right) observations with the aliases of their OBSID (see Table \ref{period}) mentioned in the figures. 
The pulse profiles are plotted in sequence of their brightness from top to bottom. Note that both the pulse peaks, especially the second peak, P2, show an evolution with luminosity.}}
\label{pp} 
\end{figure*}

\emph{Pulse profile studies:} The spin period search was carried out on the \suzaku~and \nustar~ lightcurves obtained above. Pulsations were clearly detected in nine out of ten \suzaku~ observations and five out of six observations for \nustar. The best-obtained spin period for each observation are tabulated in Table \ref{obsid} and a plot of the pulse period history is shown in Fig.~\ref{period_evolution}. There was no clear pulsation in one \suzaku~observation (60), and one \nustar~observation (06). In order to obtain the pulse profiles for these two observations, we folded the lightcurves at the period corresponding to the highest chisquare value within the period search range. Folded profile from all the observations are shown to the left (\suzaku) and right (\nustar) of Fig.~\ref{pp}. We identify the first `broader' peak as P1 and P2 as the one following it. We notice that although both the peaks evolve with luminosity, the second peak P2 decreases more rapidly with decreasing luminosity. This is especially noticeable in the \nustar~ pulse profiles on the right of Fig.~\ref{pp}. In Table \ref{obsid}, we have also listed the ratio (R) of the two peaks $P_{1}$ \& $P_{2}$ above the un-pulsed component ($P_{UP}$) defined as ($P_{1}$-$P_{UP}$)/($P_{2}$-($P_{UP}$), and the pulse fraction (P.F) defined as ($P_{max}$-$P_{UP}$)/($P_{max}$+($P_{UP}$).

\begin{table} 
\caption{Observation log for \src with \suzaku~and \nustar~ordered in sequentially as decreasing brightness states. The pulse fraction (P.F) is defined as ($P_{max}$-$P_{UP}$)/($P_{max}$+$P_{UP}$) where $P_{UP}$ is the un-pulsed component and $P_{1}$, $P_{2}$ are the two peaks. The ratio (R) is ($P_{1}$-$P_{UP}$)/($P_{2}$-$P_{UP}$).} 
\label{period}
\begin{tabular}{|c|c|c|c|c|c|c|}
  \hline
Facility~& OBSID & D.O.O & Useful exposure & P$_{spin}$& P.F & R \\ 
  & (alias in this paper) & & (ks) & (s) & \% & \\
  \hline
\suzaku~ & 706030010 (10) & 2011-04-07 & 18.5 & 0.70170860 $\pm$ 0.00000089 & 45.8 \e 1.2 & 0.99 \e 0.04 \\ 
& 706030070 (70)& 2011-09-21 & 18.1 & 0.70152930 $\pm$ 0.00000037 
& 45.9 \e 1.4 & 1.35 \e 0.06  \\
& 706030050 (50)& 2011-05-25 & 17.8 & 0.70166090 $\pm$ 0.00000016 
& 45.3 \e 1.4 & 1.10 \e 0.05 \\
& 706030080 (80)& 2011-11-10 & 19.9 & 0.70147950 $\pm$ 0.00000018 
& 40.1 \e 1.1 & 0.95 \e 0.04 \\
& 706030020 (20)& 2011-04-18 & 17.3 & 0.70169900 $\pm$ 0.00000029 
& 44.7 \e 1.3 & 1.60 \e 0.07 \\
& 706030090 (90)& 2011-12-12 & 17.3 & 0.70143900 $\pm$ 0.00000047 
& 50.6 \e 1.5 & 1.71 \e 0.09 \\
& 706030030 (30)& 2011-04-22 & 15.7 & 0.70169140 $\pm$ 0.00000065 
& 46.9 \e 1.6 & 2.02 \e 0.14 \\
& 706030100 (100)& 2012-03-19 & 18.6 & 0.70134440 $\pm$ 0.00000032 
& 46.5 \e 2.3 & 1.34 \e 0.11 \\ 
& 706030060 (60)& 2011-06-28 & 18.7 & -  
& $<$ 4.0  & - \\
& 706030040 (40)& 2011-05-10 & 17.8 & 0.70167570 $\pm$ 0.00000234 & 
34.4 \e 6.8 & 2.55 \e 1.58\\ 
\nustar~ & 30202004008 (08) & 2016-10-24 & 20.1 & 0.69953702 \e 0.00000027  
& 37.9 \e 0.2 & 0.99 \e 0.01\\
& 10002013003 (03) & 2012-08-06 & 17.0 & 0.70118470 \e 0.00000006 
& 36.3 \e 0.3 & 1.88 \e 0.03 \\
& 30202004002 (02) & 2016-09-08 & 20.8 & 0.69958702 \e 0.00000036 
& 36.1 \e 0.3 & 1.10 \e 0.01 \\
& 30202004004 (04) & 2016-09-19 & 19.9 & 0.69957772 \e 0.00000016  
& 41.3 \e 0.3 & 1.67 \e 0.02   \\
& 10002013001 (01) & 2012-07-05 & 33.1 & 0.70122190 \e 0.00000068 
& 12.2 \e 0.6 & 4.00 \e 1.38 \\
& 30202004006 (06) & 2016-10-01 & 19.4 & - 
& $<$ 1.9 & - \\
\hline
\end{tabular}
\label{obsid} 
\end{table}

\emph{Pulse period evolution:}
The pulse period evolution of \src has been investigated extensively in the past. Unlike most HMXB pulsars with supergiant companion stars which show both spin-up and spin-down episodes resulting into significant random variations in period evolution over their long term trends \citep{bildsten1997}, \src has only been found to be spinning up since its discovery. \citealt{inam2010} however, reported some variations in the spin-up rate. Starting from a value of $\sim$ 3.6$\times$10$^{-11}$ Hz s$^{-1}$, since its discovery, the spin-up rate halved to a value of $\sim$ 1.9$\times$10$^{-11}$ Hz s$^{-1}$ in about 20 years and then increased to $\sim$ 2.6$\times$10$^{-11}$ Hz s$^{-1}$ just prior to the launch of RXTE in 1995 since when there are large number of period measurements. A sudden change in the spin-up rate around MJD 50000 is also obvious from the residuals to a linear fit of the period history shown in Fig.~\ref{period_evolution}. We discuss the interpretation of this in details in section \ref{sec:disc}.

\begin{figure*}
\centering
\includegraphics[height=11.0cm,width=14.0cm,angle=0]{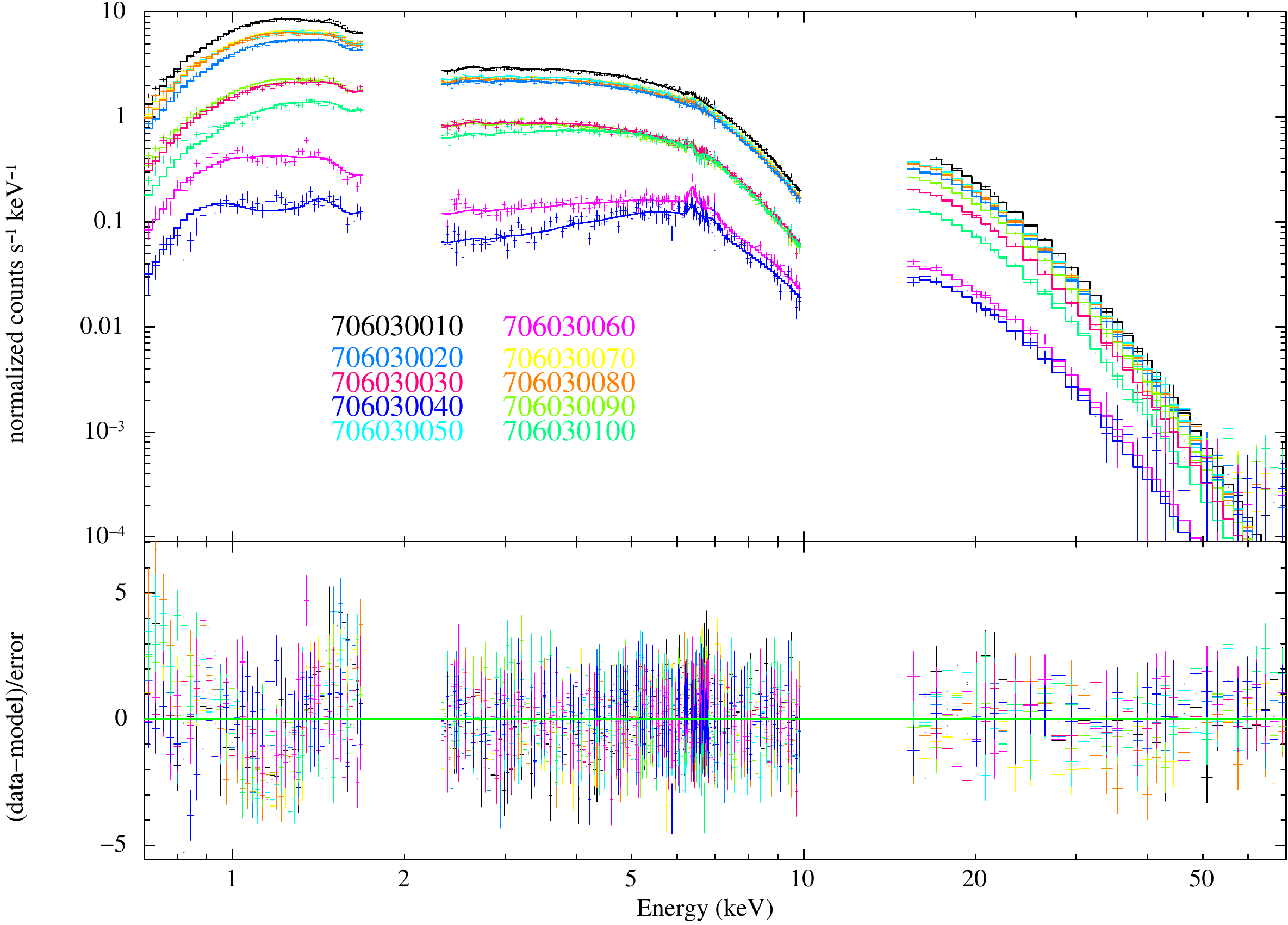}
\hspace{-6.0cm}
\includegraphics[height=11.0cm,width=14.0cm,angle=0]{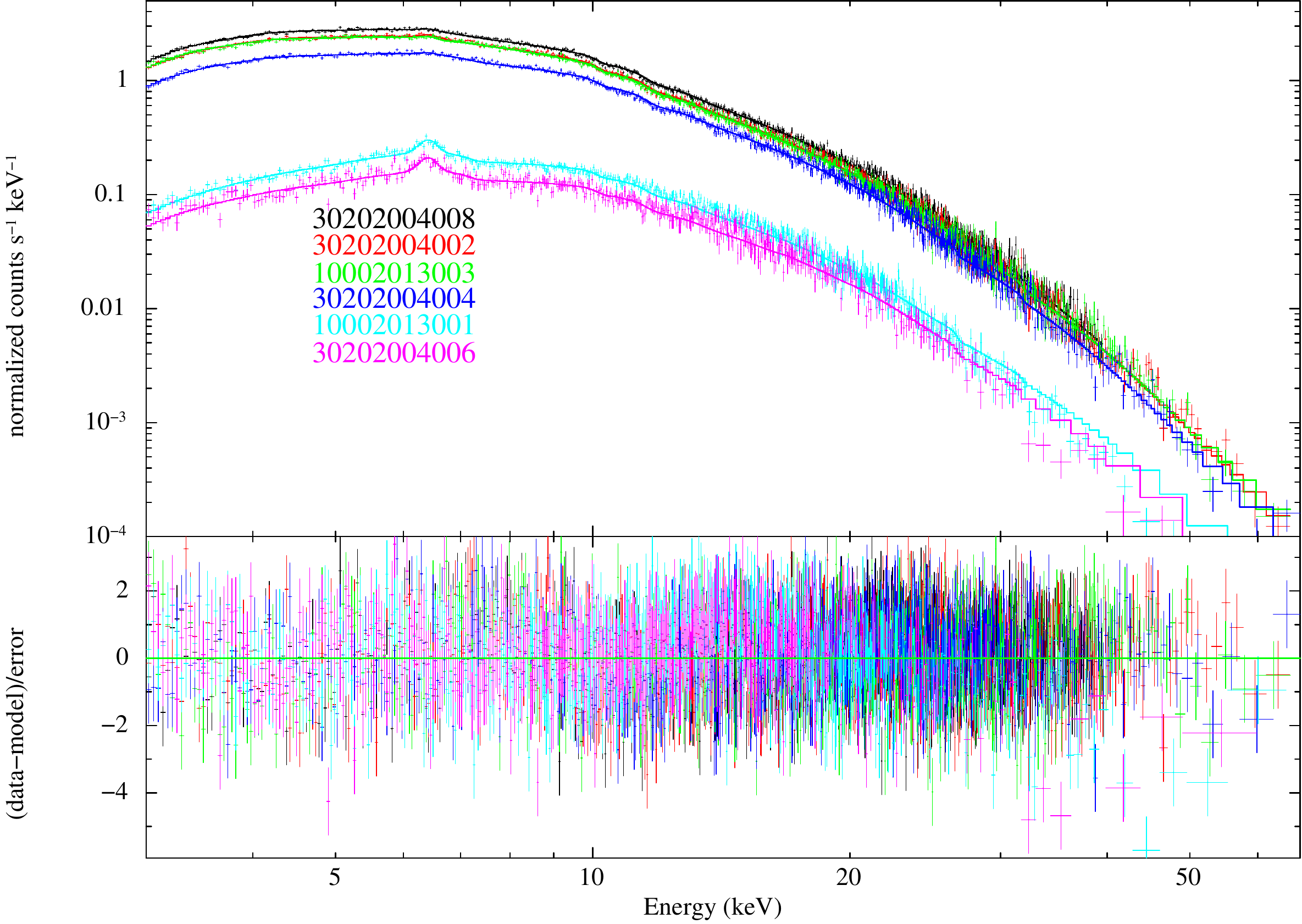}
\vspace{-0.1cm}
\caption{Top: \suzaku~spectra for all 10 observations plotted together. Below: \nustar~spectra for all 6 observations plotted together. 
Both these spectra have been fit by an absorbed powerlaw with 
exponential cutoff energy, a partial ionized absorber and a Gaussian line (plus an additional black body component for \suzaku, model 2, section 
\ref{spectral analysis}). The fit parameters of the individual fits for these observations are given in Tables \ref{suzaku_spec2} and \ref{nustar_spec}.} 
\label{spec_suzaku_nustar} 
\end{figure*}

\subsection{Spectral analysis}
\label{spectral analysis}
\subsubsection{\suzaku~}
\begin{wrapfigure}{8}{0.52\textwidth}
\centering
\includegraphics[height=10.0cm,width=10.0cm,angle=0]{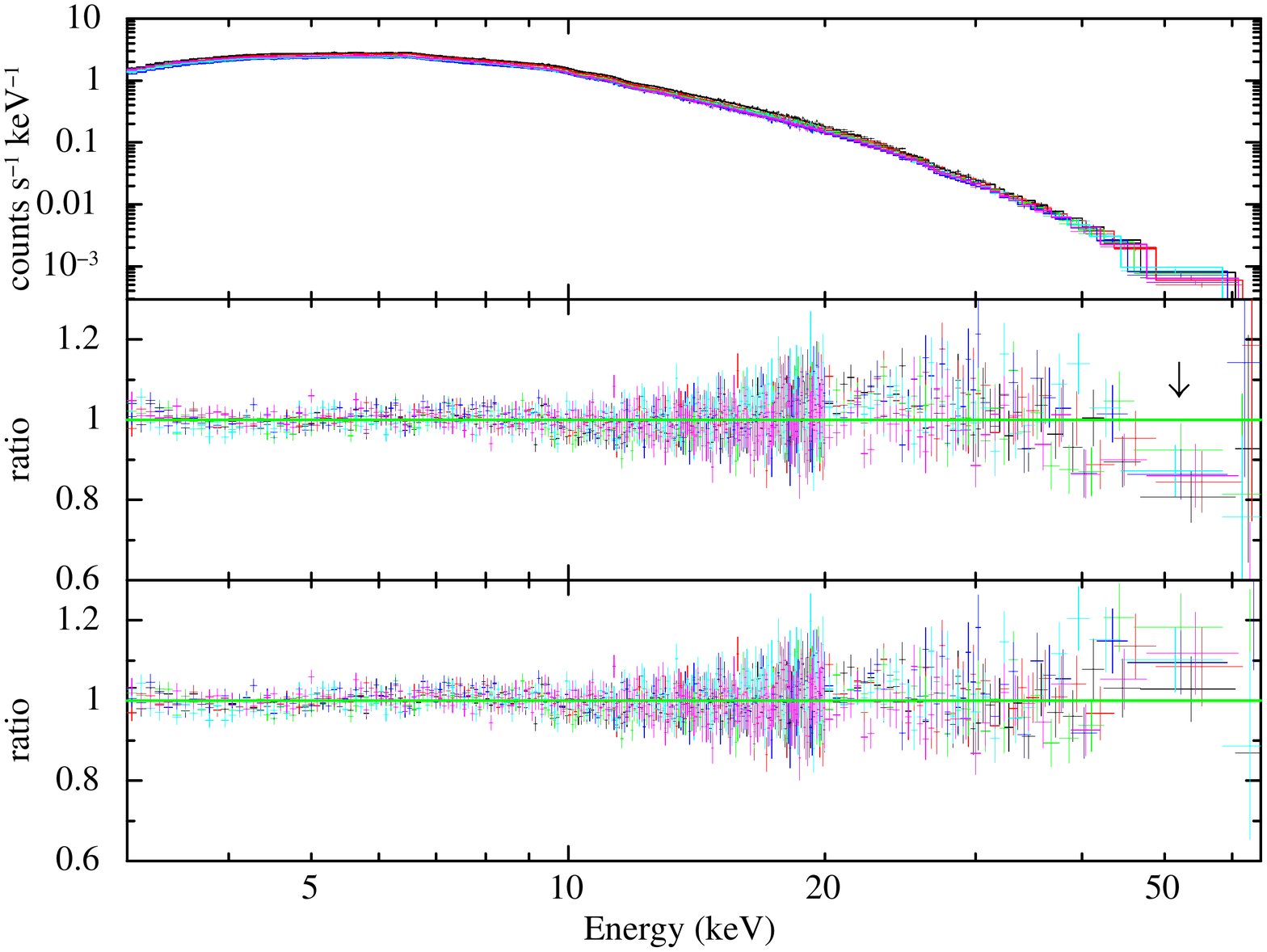}
\vspace{-0.3cm}
\caption{Plot for three \nustar~ spectra where a tentative CRSF at $\sim$ 55 keV is spotted (marked with an arrow in middle panel). The middle (lower) panel shows the ratio of data to model without (with) the CRSF model.}
\label{crsf_nustar} 

\end{wrapfigure}
We performed spectral analysis of \src using data from XIS-0 and the HXD/PIN. Spectral fitting was performed using \texttt{XSPEC} v12.10.0e. Artificial features are known in the XIS spectra around the Si edge and Au edge and the energy range of 1.75-2.23 keV is usually not used for spectral fitting. 
For each observation, we fitted the XIS and PIN spectra simultaneously with all parameters tied.
The 2048 channel XIS spectra were rebinned by a factor of 10 upto 5 \rmfamily{keV}, by 2 from 5-7 \rmfamily{keV} by 14 for the rest. The PIN spectra were binned by a factor of 2 till 22 \rmfamily{keV}, by 4 from 22-45 \rmfamily{keV}, and by 6 for the rest to ensure similar signal to noise ratio across the entire energy band. \\
We fitted each individual X-ray spectrum of \src with standard continuum models \footnote{http://heasarc.gsfc.nasa.gov/xanadu/xspec/manual/XspecModels.html} used for NS-HMXBs like an absorbed powerlaw with an exponential cutoff energy, \texttt{HIGHECUT} and \texttt{CUTOFFPL}. Both the models have been used to described the broad-band spectrum of \src in earlier works \citep[]{naik2004,pike2019}. In addition to this continuum, the spectra also necessitated a blackbody component to fit the soft excess below 2 keV as seen in many X-ray pulsars \citep[]{paul2002_smcx1,hickox2004}, as well as a Gaussian line at 6.4 keV for the K$_{\alpha}$ emission of neutral iron. The two continuum models described above fitted the X-ray spectra well with the $\chi^{2}_{r}$/d.o.f for the \suzaku~spectrum with highest photon counts (706030010) being 1.11/270, and 1.09/271 respectively. We therefore chose the \texttt{CUTOFFPL} model for further spectral analysis. In \texttt{XSPEC} notation, this model (model 1) is:
\begin{equation}
constant<1>*phabs<2>(cutoffpl<3> + bbody<4> + gaussian<5>)   
\label{model1}
\end{equation}

The complete set of parameters for the fitting above are listed in Table \ref{suzaku_spec1}. 
During the course of the spectral fitting above, we noticed that the change in the X-ray intensity of \src at different superorbital phases is not limited to the XIS energy band (0.3-10 keV) alone. The hard X-rays (15-70 keV) measured in the HXD/PIN band also change by a large factor with different superorbital states. 
This variation is of paramount importance to understand the superorbital modulation in SMC X-1. To the best of our knowledge, such a change in the hard X-ray flux at different phases of superorbital modulation has never been discussed in literature so far,  although there have been reports of changes in hard X-ray flux (Table 6 of \citealt{Brumback_2020}) at different superorbital states of SMC X-1.


In the current prevailing hypothesis, the superorbital intensity variation of SMC X-1 is caused by absorption in a precessing warped accretion disk. Such signatures are manifested as a reduction in soft X-rays (below $\sim$5 keV) in the low intensity states but the hard X-ray photons should be unaffected by this absorption. Curiously though, we notice that for \src in the low super-orbital intensity states, the reduction in soft X-rays are also accompanied by a corresponding decrease in the hard X-ray emission. Such a behaviour is in contradiction with absorption caused by a neutral absorber. A further signature of absorption in neutral medium is the fact that the spectral shape, especially below 3 keV, have a strong energy dependence in soft X-rays (see various examples in Fig.~A.1 of \citealt{pradhan2018}). The XIS spectra of the \suzaku~observations of SMC X-1 however do not show any such dependence characteristic of neutral absorption (Fig.~\ref{spec_suzaku_nustar} top). Motivated by this, we therefore introduced a partially ionized absorber to the above model (say, model 2) which in \texttt{XSPEC} notation is given by the equation below:
\begin{equation}
constant<1>*phabs<2>(zxipcf<3>*cutoffpl<4> + bbody<5> + gaussian<6>)
\label{model2}
\end{equation}
The use of ionized absorber is also justified since such high X-ray luminosity as observed in \src contribute to ionization of matter around the neutron star. This is also supported by the remarkably low value of equivalent width of the neutral iron K$_{\alpha}$ line of \src as compared to other similar systems like LMC X-4. The complete set of parameters for the fitting above are listed in Table \ref{suzaku_spec2} and individual spectra fit with this model are plotted together on top of Fig.~\ref{spec_suzaku_nustar}.

\subsubsection{\nustar}
We fitted the \nustar~FPMA and FPMB spectra simultaneously with all parameters tied, except the relative instrument normalizations of which were kept free. The 4096 channel FPM spectra were rebinned by a factor of 2 upto 20 \rmfamily{keV}, by 10 from 20-60 \rmfamily{keV} by 6 for 60-78\,keV and by a factor of 2 for the rest. \\
We use the 3-70\,keV \nustar~spectra for spectral fitting using with the same continuum model with partially ionized absorber as \suzaku. The only difference being that we did not require any blackbody component to fit the soft excess in contrast to the \nustar~analysis by \citet{pike2019}. (With the sensitivity of \nustar~ below 3\,keV being very poor, a blackbody component with a temperature of $\sim$ 0.16 keV cannot be constrained with \nustar). 
The fits for all the observations are tabulated in Table \ref{nustar_spec} and the individual spectra are plotted on the bottom of Fig.~\ref{spec_suzaku_nustar}. 
We also noted an absorption feature in the spectrum reminiscent of a  Cyclotron Resonant Scattering Feature (CRSF) around 50-60\,keV in the three brightest observations 
(08, 02, 03) with \nustar. We used \texttt{GABS} to model this cyclotron absorption feature that decreased the  $\chi^{2}$/d.o.f from 822/701 to 751/699, 845/700 to 732/698 and 842/700 to 750/698 for observation 08, 02, 03 respectively. The ratio of data to model for these three observations with and without the CRSF component is shown in Fig.~\ref{crsf_nustar}.

\subsection{Joint spectral fitting for different superorbital states}
\label{joint}
\subsubsection{\suzaku} 
In the prevailing hypothesis for superorbital intensity variation in \src (also LMC X-4), the variation is caused by absorption in a precessing warped accretion disk while the intrinsic luminosity of the source remains nearly constant as seen in the soft, $\sim$ 0.1-10\,keV X-ray spectrum (see, e.g., \citealt{wojdowski1998}). The nearly constant peak intensity of the \spo modulation observed for more than two decades with RXTE-ASM (1.5-12.0 keV band, from 1996-2011), Swift-XRT and MAXI-GSC supports this hypothesis. All of these observations are however well below 20\,keV. The aim of this current work is to investigate if the multiple broad band spectra also show evidence of variable absorption.
\begin{figure}
\centering
\includegraphics[height=12.0cm,width=9.0cm,angle=0]{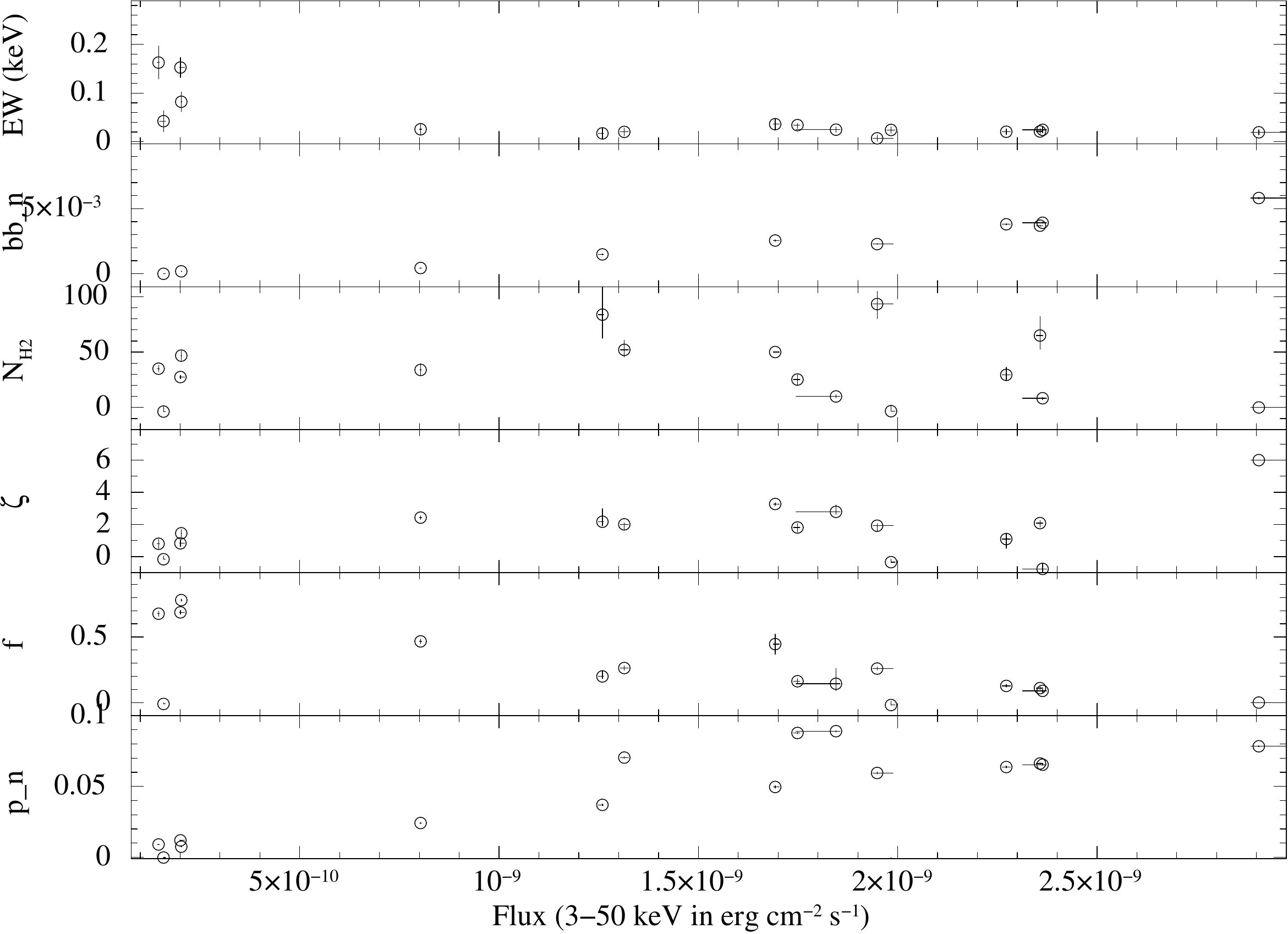}
\caption{Model parameters versus the 3-50 keV flux corresponding to these 16 observations for joint fits (10 
\suzaku~spectra fit jointly and 6 \nustar~spectra fit jointly) using a powerlaw and an exponential cutoff with a partially ionized absorber, a Gaussian line for neutral iron K$_{\alpha}$ line (and black-body component for \suzaku~spectra only). For this joint fitting, we allow the black-body, powerlaw normalization, 
absorption parameters (f, \nhh, $\xi$) and Gaussian normalization to vary. See section \ref{joint} for details. }
\label{spec_joint} 
\end{figure}

To achieve this, we performed a joint spectral fitting of the 10 \suzaku~observations. We first fit the continuum of the brightest observation (10) from both XIS and PIN with an absorbed powerlaw and high energy cutoff coupled with an ionized partial absorber. We then tied all the spectra from other \suzaku~observations to this model ($N_{\rm H1}$ was frozen at 0.3 $\times$ $10^{22}$ atoms cm$^{-2}$) for simultaneous fitting. 
With this continuum (10) as the reference, we therefore performed joint fitting of the spectra obtained in other intensity states while allowing the absorption parameters, i.e., covering fraction (f), local absorption column density (\nhh) and ionization parameter $\xi$ to change. 

In case of \suzaku~where we could constrain the soft excess component, we allowed the black body normalization to vary. Furthermore, we allowed the Gaussian normalization of the 6.4\,keV line to change as well. With the above procedure, the low energy part of the spectrum fit considerably well with the low-intensity states exhibiting a higher value of covering fraction. However, the systematic residuals above 10\,keV exhibited a misfit in the hard X-ray band. 
We therefore freed the powerlaw normalization of the 10 spectra which drastically improved the fits. The powerlaw normalization differ by as much as a factor of 6. Seeing such a large change in this powerlaw component, we cross-checked the peak-to-peak superorbital variation in the long-term BAT lightcurves. Interestingly, we find that the peak-to-peak variation of the superorbital modulation in the long-term BAT lightcurves is only $\sim$ 15\%. Such a contrast in the powerlaw normalization versus the orbit averaged BAT lightcurves is puzzling and we will discuss this further in section \ref{sec:disc}.


\subsubsection{\nustar}
For joint fitting of 6 \nustar~observations, we follow the same methodology as above with the same model (without the black body component which cannot be constrained with \nustar). We fit the spectrum with highest statistics (08) and tied other 5 spectra with this, while allowing the absorption parameters (f, \nhh, $\xi$), powerlaw and Gaussian normalization to be free. We also find the same results as obtained with \suzaku~ and the powerlaw normalization vary by a factor of 6. The physical interpretation of this exercise will be explained in section \ref{sec:disc}. \\

This variation of the spectral parameters in both the exercises above as a function of X-ray flux are shown in Fig.~\ref{spec_joint}. 

\section{Discussion}
\label{sec:disc}
Individual broad band spectra of all observations can be fitted with a \texttt{CUTOFFPL}, along with a blackbody for a soft excess (only \suzaku) and neutral iron K$\alpha$ line. Equivalent width of the iron line vary from 10-270 eV, anti-correlated with luminosity. Through joint fitting of the 16 spectra in different superorbital states, we find that the powerlaw normalization vary by a factor of upto 6 while the overall variation of the peak of the superorbital modulation obtained from the BAT lightcurves is $\sim$ 15\%. In the event of this X-ray variation caused only through absorption in precessing warped accretion disk, the power-law normalisation should not vary beyond the variation in the peak luminosity of the superorbital variation (i.e., upto 15\% in case of SMC X-1). This huge difference in the percentage of variation in BAT lightcurves versus powerlaw normalization can only be explained if we introduce an intrinsic variability of X-ray emission from the source as opposed to superorbital modulation caused by absorption\footnote{Note that taking a cue from the shape of XIS spectrum, the large luminosity of SMC X-1 that ionizes matter, and a very weak iron K$\alpha$ line, we have used an ionized absorber as opposed to neutral.}


Pulsations are detected in all but two observations of \suzaku~(15-70\,keV)~and \nustar~(3-70\,keV). We also note that the pulse fraction in hard X-rays ($>$ 12 keV ) remain almost constant for 13 out of 16 observations (Table \ref{period}). Considerable luminosity dependence is seen in the pulse profile, ratio of the second to the first peaks becomes smaller at lower intensities. This variation of pulse profile morphology also support our hypothesis that there is an intrinsic change in the source spectrum of \src with varying super-orbital states.

The timing analysis also reveal that interestingly (except for one \suzaku~and one \nustar~observation), the lower intensity states in \suzaku~are also pulsed in hard X-rays. This suggests that perhaps most X-rays travel across the warp instead of being scattered which would otherwise have caused smearing of pulsations. Detection of strong pulsations and relatively low EW of iron line compared to other similar systems indicate that even in the low state we are mostly seeing the central object directly and not in scattered radiation. 

From the residuals to a linear fit of the period history shown in Fig.~\ref{period_evolution}, it is obvious that there was a sudden change in spin-up rate around MJD 50000 (year 1995). This abrupt change was also noted from some ROSAT observations, though it was not specifically mentioned in the report \citep{kahabka1999}. In the scenario that the superorbital intensity variation of \src is due to variable absorption, the accretion torque onto the neutron star should be compared against the peak luminosity of the superorbital variation. We therefore compare the peak luminosity before and after the sudden change in spin-up rate (around MJD 50,000). The long term flux variation of \src was investigated in detail with HEAO 1 for three cycles and they reported a peak flux of 2.8$\times$10$^{-11}$ erg cm$^{-2}$ s$^{-1}$ in the 13-70 keV band \citep{gruber1984}. In the current observations with \suzaku~and \nustar, some of which are near the peak of the superorbital modulation, the flux in the same energy band is measured to be much higher (Tables \ref{suzaku_spec2} and \ref{nustar_spec}), upto a factor 50 compared to the HEAO 1 measurements. It is therefore likely that there was a significant change in the mass accretion rate onto SMC X-1 associated with the change in spin-up rate some time around around MJD 50000 (year 1995). Subsequent to this event, there was a decrease in the superorbital period of SMC X-1 from about 60 days to about 45 days and a weak correlation was found between the superorbital period and the short term spin-up rate \citep{dage2019}, which may therefore be related to the change in mass-accretion rate. This finding therefore directly support our hypothesis of varying X-ray emission from the source.

It is worthwhile to clarify here that our findings do not question the {\it presence} of a PWAD in SMC X-1, but rather shows that the {\it absorption} in the PWAD is not the cause of \spo modulation - at least not wholly - and there are signatures of intrinsic changes in X-rays emitted from the neutron star. To reiterate, in the PWAD model, the warped accretion disk is irradiated by the X-rays from the pulsar beam where this hard X-ray photons get reprocessed to soft X-rays. Therefore, by studying the differences in pulse profiles of the direct (hard) and the reprocessed (soft) X-ray emission, constraints can be placed on the disk geometry \citep{hickox2005}. Recently, \citealt{Brumback_2020} used simultaneous \xmm and \nustar~data that span a complete superorbital cycle to illustrate this model. They find that long term changes in soft pulse shape and phase are consistent with reprocessed emission from a precessing inner disk for two (08 and 02 in this paper) out of four observations. In this work too, we note that the variation in soft thermal component clearly map the hard X-rays (black-body versus power-law normalization in Fig.~\ref{spec_joint}) thereby indicating that the hard X-rays are reprocessed, possibly in the accretion disk. 

We should also mention here that in observation 04, the same authors note a change in the shape of the \nustar~pulse profiles. Such a change in shape of pulse profiles for obs 04 is also noted in our work (right of Fig.~\ref{pp}) and is one of the evidences that indeed there is some intrinsic change in the hard X-ray emission from the central object. Finally, the authors find no pulsations in observation 06 similar to what is reported here. Such an extinction of pulses (in first half of observation 01) has been explained by \citealt{pike2019} as being caused by obscuration by a Compton thick matter in the accretion disk or inhibition of accretion caused by the onset of propeller regime as seen in the case of Ultra Luminous X-ray sources (ULXs). The latter is ruled out since it is characterized by dramatic flux variability as opposed to `continuous' transition between high and low states like seen in SMC X-1. While the former is still a possibility when we consider scattering of the soft X-rays, it is unlikely that the hard X-rays from the neutron star are scattered enough to switch-off pulsations altogether. 

The broad band timing and spectral characteristics of \src as reported here from ten \suzaku~and six \nustar~observations indicate that a variable accretion rate is the possible reason behind the superorbital intensity variation in SMC X-1. Since the X-ray luminosity drives the warp, the configuration of the warp should be sensitive to the mass transfer rate onto the compact object. This possibility of the variable accretion rate is what we have discussed throughly in this paper and have provide many spectral and timing characteristics to solidify our claim that the superorbital modulation is not due to absorption in the precessing warped disk \emph{alone}.

One possible explanation for \spo modulation in SMC X-1 drawn analogously from the studies of cataclysmic variables (CVs) is the generation of `bright spots'. If the mass transfer rate from the donor star is quasi-steady, the local values of mass transfer in accretion discs can exceed the actual mass transfer rate from the companion star \citep{rutten1992}. In this scenario, there is a creation of `bright spots' in the accretion disk  and the warp in \src may transport this mass transfer stream close to the neutron star.  As the warp precesses, this bright spot move through the disc, thus varying the brightness and altering the X-ray spectrum at different \spo phases like seen here. Such a possibility although mentioned by earlier authors in literature (Discussion in \citealt{clarkson2003_smcx1}) suffer from a number of drawbacks. In particular, there is no reason to believe that the amount of energy emitted by these bright spots will be a significant when compared to X-rays emitted from the neutron star. Additionally, these bright spots, if present, will be formed well above the co-rotation radius of the accretion disk and is not expected to be pulsed. 

In order to gain further insights into the variation of \spo in SMC X-1, there needs to be further in-depth analysis by studying line emissions using high resolution spectra. The variability of these emission lines can constrain the geometry and dynamics of the line-emitting regions in such sources (e.g., LMC X-4; \citealt{neilsen2009}). For instance, visibility of some Doppler-shifted lines with \spo phases is a direct evidence for the precession of accretion disk. During low (high) \spo states when the disk is viewed almost edge-on (face-on), the Doppler-shifted lines originating in the inner accretion disk appear (disappear). On the other hand, if the strength of emission lines do not change with \spo phases, the emission lines probably originate in region that subtend a large solid angle to the compact object, possibly the stellar wind (or outer disk). This way, by studying the variability of line fluxes and energies of emission lines we can investigate the physical origin of such \spo modulation in SMC X-1. High-resolution spectra can also be used to investigate if the \spo modulation is caused by vertical columns in the  accretion disk (e.g., EXO 0748-676; \citealt{garate_exo0748}). Such analysis are beyond the scope of this paper here and we leave that for future works.

\section{Summary}
\label{sec: conc}
We summarize below the main findings of this work:
\begin{itemize}

\item We report broad-band spectral-timing results using sixteen observations of \src with \suzaku~and \nustar. All individual broad band 
spectra can be fitted with an absorbed high energy cutoff powerlaw continuum along with a soft blackbody and a weak iron emission line (model 1). Another 
spectral model with absorbed high energy cutoff powerlaw and a partially ionized absorber (model 2) hint an increase in covering fraction with decreasing
intensity of the source. The line equivalent varies from ~10 eV in high state to $\sim$ 270 eV in low state.

\item Joint spectral fits using model 2 (without black-body component for \nustar) from all intensity states cannot be explained as being caused by
absorption of an otherwise stable source. The change in the normalization of the power-law (factor of $\sim$ 6 in the joint fits) is much larger than expected 
from changes in the superorbital intensity modulation due to absorption the precessing disk alone.  An alternative possibility is the change in the X-ray emission 
from the source directly that cause such variability. 

\item Pulsations are detected in all but one \suzaku~and one \nustar~ observation. The detection of pulsation in most of the low-intensity state observations 
indicate a direct view of the compact object event in the low states in contrary to the prevailing belief of the superorbital modulation being caused by absorption 
in precessing accretion disk alone. 
    
\item The pulse profile shows intensity dependence, the second peak becoming less prominent at lower intensities. These changes in the pulse profiles indicate an 
intrinsic change in the beaming pattern with the intensity states which could be connected to a change in the accretion rate.

\item A putative CRSF is detected at $\sim$ 55 keV in the the brightest \nustar~observations indicates a surface magnetic field of $\sim$ 4.2 $\times$ 10$^{12}$ G for \src.

\item Period history of \src is extended by about 13 years, continues to spin-up. It shows a sudden change in the spin-up rate, perhaps along with a large change in peak luminosity.

\end{itemize}

\section*{Acknowledgement}
The authors would like to thank the reviewer for his/her contributions to the paper in the form of useful comments and suggestions 
that drastically improved the quality of the paper. This research has made use of data and software provided by the High Energy Astrophysics Science 
Archive Research Center (HEASARC), which is a service of the Astrophysics Science Division at NASA/GSFC.

\begin{table}
\renewcommand{\tabcolsep}{0.8pt}
\scriptsize
 \caption{Best-fit parameters of \src during \suzaku~with model 1 (eqn. \ref{model1}, see section \ref{spectral analysis} for details). Errors quoted are for 90 per cent confidence range. The observations are arranged in the order of decreasing brightness states with high, medium and low states designated as H, M and L respectively.}
 \begin{tabular}{|c|c|c|c|c|c|c|c|c|c|c|}
\hline
  & \multicolumn{9}{c}{Obs} & \\
\hline
 & H & H & H & H & H & H & M & M & L & L\\
& 10 & 70 & 50 & 80 & 20 & 90 & 30 & 100 & 60 & 40 \\
\hline
$N_{H1}^{a}$ & 0.15 $\pm$ 0.02& 0.15 \e 0.02 & 0.20 \e 0.02  & 0.09 \e 0.02 & 0.22 \e 0.03 & 0.11 \e 0.04 & 0.16 \e 0.04 & 0.14 \e 0.06 & 0.04 \e 0.04 & 0.09 \e 0.09 \\ 

$\Gamma$ & 0.48 $\pm$ 0.02 & 0.49 \e 0.03 & 0.43 \e 0.03  & 0.43 \e 0.02 & 0.50 \e 0.03  & 0.57 \e 0.03 & 0.49 \e 0.04 & 0.22 \e 0.04 & -0.42 \e 0.08 & -1.05 \e 0.11\\ 

E$_{f}$ &  9.4 $\pm$ 0.3 & 9.5 \e 0.3 & 9.2 \e 0.3  & 9.2 \e 0.3 & 10.8 \e 0.3 & 10.6 \e 0.3 & 9.9 \e 0.4 & 8.2 \e 0.4 & 7.9 \e 0.5 & 5.7 \e 0.4\\ 

$\Gamma_{norm}^{b}$ & 0.067 \e 0.002  & 0.052 \e 0.001 & 0.050 \e 0.001  & 0.046 \e 0.001 & 0.038 \e 0.001 & 0.039 \e 0.002 & 0.026 \e 0.001 & 0.011 \e 0.001 & 0.0005 \e 0.0001 & 0.0002 \e 0.00001\\ 

$kT$ &  0.22 $\pm$ 0.008 & 0.22 \e 0.008 &  0.21 \e 0.007  & 0.23 \e 0.01 & 0.21 \e 0.01 & 0.22 \e 0.02 & 0.21 \e 0.01 & 0.17 \e 0.01 & 0.23 \e 0.02 & 0.25 \e 0.03 \\ 
                                        
$kT_{norm}^{c}$ & 2.8 $\pm$ 0.3 & 1.7 \e 0.2 & 2.4 \e 0.3  & 1.4 \e 0.2 & 1.4 \e 0.2 & 0.9 \e 0.2 & 0.7 \e 0.2 & 0.3 \e 0.1 & 0.09 \e 0.02 & 0.03 \e 0.01 \\ 
  
K$\alpha$ & 6.38 $\pm$ 0.05 & 6.45 \e 0.04 & 6.43 \e 0.05  & 6.36 \e 0.06 & 6.49 \e 0.07 & 6.36 \e 0.03 & 6.27 \e 0.08 & 6.30 \e 0.03 & 6.41 \e 0.02 & 6.35 \e 0.05 \\ 
                                        
EW & 0.015 \e 0.005 & 0.026 \e 0.007 & 0.022 \e 0.005  & 0.022 \e 0.006 & 0.013 \e 0.006 & 0.038 \e 0.011 & 0.024 \e 0.011 & 0.038 \e 0.011 & 0.135 \e 0.022 & 0.094 \e 0.022\\ 

Flux$^{d}$ & 2.96 $\pm$ 0.05 & 2.70 \e 0.04 & 2.58 \e 0.02  & 2.32 \e 0.04 & 1.99 \e 0.02 & 1.73 \e 0.01 & 1.31 \e 0.02 & 0.67 \e 0.02 & 0.27 \e 0.01 & 0.15 \e 0.01\\ 

$\chi^{2}_{r}$/d.o.f & 1.09/271 & 1.05/270 & 1.3/270  & 1.09/270 & 1.42/271 & 1.00/271 & 1.07/270& 1.36/270 & 1.59/271 & 1.23/271\\ 

\hline
\end{tabular}
\label{suzaku_spec1} \\
$^a$ In units of $10^{22}$ atoms cm$^{-2}$ \\
$^b$  In units of photons/keV/cm$^2$/s at 1 keV \\
$^c$ In units of L$_{39}$/$D_{10}$ where  L$_{39}$ is source luminosity in units of $10^{39}$ erg s$^{-1}$ and $D_{10}$ is the source distance in 10 kpc units. \\
$^d$ In 1-70 keV, in units of 1$\times$10$^{-9}$ erg cm$^{-2}$ s$^{-1}$  
\end{table}

\begin{table}
\scriptsize
\renewcommand{\tabcolsep}{0.7pt}
 \caption{Best-fit parameters of \src during \suzaku~with model 2 (eqn. \ref{model2}, see section \ref{spectral analysis} for details). The observations are arranged in the order of decreasing flux states, High (H), medium (M) and low (L) respectively.}
\begin{tabular}{|c|c|c|c|c|c|c|c|c|c|c|}
\hline
 & \multicolumn{9}{c}{Obs} & \\
\hline
& H & H & H & H & H & H & M & M & L & L \\
& 10 & 70 & 50 & 80 & 20 & 90 & 30 & 100 & 60 & 40 \\
\hline
$N_{H1}^{a}$ & 0.17 $\pm$ 0.01 & 0.17 \e 0.03 & 0.24 \e 0.03    & 0.14 \e 0.02 & 0.26 \e 0.02 & 0.15 \e 0.04 & 0.17 \e 0.06 & 0.15 \e 0.10 & 0.21 \e 0.07 & 0.17 \e 0.02 \\

$N_{H2}^{a}$ & 23 $\pm$ 8 & 30 \e 2 & 48 \e 12  & 23 \e 8 & 95 \e 11 & 23 \e 10  & 81 \e 9 & 32 \e 9 & 41 \e 11 & 25 \e 7\\

$\xi$ & 1.87 \e 0.66 & 1.90 \e 0.19 & 1.89 \e 0.32  & 1.77 \e 0.34 & 2.04 \e 0.14 & 1.85 (-2.92, 0.96) & 2.14 \e 0.22 & 3.06 \e 0.17 & 1.19 \e 0.36 & 0.85 \e 0.5  \\

$f$ & 0.10 \e 0.02 & 0.18 \e 0.07 & 0.25 \e 0.03  & 0.09 \e 0.01  & 0.27 \e 0.01 & 0.03 (-0.01, 0.02) & 0.27 \e 0.11 & 0.44 \e 0.07 & 0.75 \e 0.04 & 0.82 \e 0.09\\

$\Gamma$ & 0.59 $\pm$ 0.02 & 0.65 \e 0.01 & 0.60 \e 0.02  & 0.51 \e 0.02 & 0.60 \e 0.02 & 0.59 \e 0.03 & 0.51 \e 0.03 & 0.59 \e 0.03 & 0.59 \e 0.07 & 0.69 \e 0.09 \\

E$_{f}$ &  9.7 $\pm$ 0.2 & 10.1 \e 0.3 & 9.6 \e 0.2  & 9.6 \e 0.3 & 10.4 \e 0.3 & 9.6 \e 0.3 & 9.5 \e 0.62 & 9.8 \e 0.4 & 10.1 \e 0.9 & 12 \e 5\\

$\Gamma_{norm}^{b}$ & 0.091 $\pm$ 0.005 & 0.069  \e 0.01 & 0.078 \e 0.004  & 0.049 \e 0.004 & 0.063 \e 0.008 & 0.036 \e 0.006 & 0.033 \e 0.001 & 0.015 \e 0.001 & 0.006 \e 0.003 & 0.003 \e 0.001\\

$kT$ &  0.21 $\pm$ 0.002 & 0.21 \e 0.01 & 0.19 $\pm$ 0.01  & 0.20 \e 0.01 &  0.19 \e 0.01 & 0.19 \e 0.01 & 0.21 \e 0.02 & 0.16 \e 0.03 & 0.17 \e 0.01 & 0.17 \e 0.05 \\
                                        
$kT_{norm}^{c}$ & 2.92 $\pm$ 0.16 & 1.79 \e 0.26 & 2.79 $\pm$ 0.01  & 1.67 \e 0.24 & 1.71 \e 0.30 & 1.15 \e 0.29 & 0.77 \e 0.21 & 0.28 \e 0.32 & 0.17 \e 0.07 & 0.05 \e 0.09 \\
 
K$\alpha$ & 6.37 $\pm$ 0.04 & 6.46 $\pm$ 0.06 & 6.41 $\pm$ 0.06  & 6.39 $\pm$ 0.06 & 6.52 \e 0.12 & 6.38 \e 0.04 & 6.28 \e 0.10 & 6.32 \e 0.04 & 6.41 \e 0.03 & 6.35 \e 0.06 \\
                                        
EW & 0.014 $\pm$ 0.007 & 0.024 $\pm$ 0.006 & 0.013 $\pm$ 0.007  & 0.020 $\pm$ 0.002 & 0.011 \e 0.006 & 0.040 $\pm$ 0.011 & 0.015 $\pm$ 0.011 & 0.035 $\pm$ 0.012 & 0.100 $\pm$ 0.017 & 0.053 $\pm$ 0.025 \\

Flux$^{d}$ & 3.62 $\pm$ 0.06 & 2.70 $\pm$ 0.07 & 2.54 $\pm$ 0.06  & 2.51 $\pm$ 0.03 & 2.21 $\pm$ 0.05 & 1.94 $\pm$ 0.05 & 1.43 $\pm$ 0.04 & 0.78 $\pm$ 0.02 & 0.22 $\pm$ 0.01 & 0.17 $\pm$ 0.09\\

Flux$^{e}$ & 1.44 \e 0.01 & 1.21 \e 0.01 & 1.18 \e 0.01  & 1.16 \e 0.01 & 1.06 \e 0.01 & 0.89 \e 0.01 & 0.66 \e 0.03 & 0.43 \e 0.02 & 0.13 \e 0.10 & 0.10 \e 0.10 \\

$\chi^{2}_{r}$/d.o.f & 0.81/267 & 0.96/267 & 1.22/267  & 1.00/267 & 1.25/268 & 0.89/268 & 1.0/268 & 1.14/267 & 1.35/267 & 1.00/268 \\

\hline
\end{tabular}
\label{suzaku_spec2} \\
$^a$ In units of $10^{22}$ atoms cm$^{-2}$ \\
$^b$  In units of photons/keV/cm$^2$/s at 1 keV \\
$^c$ In units of 1$\times$10$^{-3}$ L$_{39}$/$D_{10}$ where  L$_{39}$ is source luminosity in units of $10^{39}$ erg s$^{-1}$ and $D_{10}$ is the source distance in 10 kpc units. \\
$^d$ In 1-70 keV, in units of 1$\times$10$^{-9}$ erg cm$^{-2}$ s$^{-1}$ \\
$^e$ In 13-70 keV, in units of 1$\times$10$^{-9}$ erg cm$^{-2}$ s$^{-1}$  
\end{table}

\begin{table*}
\centering
\scriptsize
 \caption{Best-fit parameters of \src during \nustar~observations with model 2 minus the black-body component 
 (see text in section \ref{spectral analysis} for details). Errors quoted are for 90 per cent confidence range. The observations are separated into three states H, M, L for high, medium and low states respectively.}
\begin{tabular}{|c|c|c|c|c|c|c|c|}
\hline
  & \multicolumn{5}{c}{OBSID} & \\
\hline
& H & H & H & M & L & L \\
 & 30202004008 & 10002013003  & 30202004002 & 30202004004 & 10002013001 & 30202004006 \\
\hline

$N_{H1}^{a}$ & 0.40 \e 0.11 & 0.46 & 0.46 & 0.46 & 2.0 \e 0.15 & 4.3 \e 0.2  \\ 

$N_{H2}^{a}$ & 22 \e 2 & 23  & 11    & 43  & 51 \e 0.77 & 85 \e 2\\ 

$\xi$ & 1.36 \e 0.30 & 2.52 \e 0.07 & 0.19 \e 0.30  & 1.22 \e 0.17 & 1.21 \e 0.10 & 0.71 \e 0.11 \\

$f$ & 0.22 \e 0.05 & 0.13 \e 0.01 & 0.17 \e 0.01  & 0.34 \e 0.01 & 0.75 \e 0.03 & 0.73 \e 0.01\\

$\Gamma$ & 0.86 \e 0.01 & 0.85 \e 0.01 & 0.85 \e 0.01 & 0.85 \e 0.01 & 0.84 \e 0.02  & 0.85 \e 0.01\\

E$_{f}$ & 11.1 \e 0.1 & 11.05 \e 0.12  & 11.65 \e 0.11  & 10.98 \e 0.08 & 8.69 \e 0.06 & 7.9 \e 0.4 \\

$PL_{norm}^{b}$ & 0.119 \e 0.002 &  0.104 \e 0.005 & 0.091 \e 0.010  & 0.077 \e 0.001 & 0.023 \e 0.002 & 0.021 \e 0.004\\

K$\alpha$ & 6.44 \e 0.06 & 6.50 \e 0.05 & 6.42 \e 0.07  & 6.56 \e 0.07 & 6.38 \e 0.03 & 6.32 \e 0.04 \\

EW & 0.023 \e 0.005 & 0.030 \e 0.007 & 0.105 \e 0.012 \e  & 0.07 \e 0.01 & 0.170 \e 0.015 & 0.269 \e 0.114\\

E$_{C}$ & 52.8 \e 1.9 & 51.53 \e 1.15 & 53.44 \e 1.30  & - & - & -\\

$\sigma$ & 9.14 \e 1.05 & 9.77 \e 0.97 & 11.11 \e 1.1  & - & - & -\\

$\tau$ & 10 & 11.87 & 16.39 \e 7.0  & - & - & -\\

Flux$^{c}$ (1-70 keV) & 2.04 \e 0.03 & 2.07 \e 0.02 & 1.81 \e 0.01  & 1.36 \e 0.01 & 0.19 \e 0.02 & 0.140 \e 0.010\\

Flux $^{c}$ (13-70 keV) & 0.92 \e 0.02 & 0.85 \e 0.01 & 0.81 \e 0.07  & 0.64 \e 0.05 & 0.11 \e 0.02 & 0.08 \e 0.21 \\

$\chi^{2}_{r}$/d.o.f & 1.07/697 & 1.10/696 & 1.05/698  & - & - & - \\

(without GABS) & 1.19/699 & 1.20/700 & 1.21/700  & 1.35/699 & 1.02/700 & 1.02/699\\
\hline
\end{tabular}

$^a$ In units of $10^{22}$ atoms cm$^{-2}$ \\
$^b$  In units of photons/keV/cm$^2$/s at 1 keV \\
$^c$ In units of 1$\times$10$^{-9}$ erg cm$^{-2}$ s$^{-1}$  

\label{nustar_spec}
\end{table*}

\bibliographystyle{pwterse-ay}
\bibliography{reference_smc}

\end{document}